\documentclass[aps,pra,twocolumn,a4paper]{revtex4-1}
\usepackage{amsmath}
\usepackage{braket}
\usepackage[pdftex]{graphicx}

\begin{document}
\title{Single Spin Detection with Entangled States}
\author{Hideaki Hakoshima}\email{hakoshima-hideaki@aist.go.jp} \author{Yuichiro Matsuzaki}\email{matsuzaki.yuichiro@aist.go.jp} 
\affiliation{
Nanoelectronics Research Institute
National Institute of Advanced Industrial Science and Technology (AIST)\\
1-1-1 Umezono, Tsukuba, Ibaraki 305-8568, Japan
}
\begin{abstract}
Single spin detection is one of the important tasks in the field of quantum metrology.
Many experiments about the single spin detection has been performed. However, due to the weak magnetic fields from the single spin, a long measurement time is required to achieve a reasonably high signal-to-noise ratio.
Here, we propose an alternative way to realize rapid and accurate single spin detection with entangled states. 
While it is known that entanglement can improve the sensitivity to measure globally applied magnetic fields,
we investigate a strategy to use the entanglement for detecting spatially inhomogeneous magnetic fields from the target single spin.
We 
show that the entanglement significantly increases the signal to noise ratio for the single spin detection even under the effect of realistic noise. Our results pave the way for practical  single spin detection.
\end{abstract}

\maketitle

\section{Introduction}
Magnetic field sensors with high sensitivity have been studied for many years due to the 
wide applications
in various areas, such as biology and materials sciences.
For the detection of the weak magnetic fields, 
many types of 
magnetic field sensors have been developed
 such as SQUID \cite{jaklevic1965macroscopic,fagaly2006superconducting,fagaly2006superconducting}, Hall sensors \cite{howells1999scanning}, 
and optomechanics \cite{forstner2014ultrasensitive}.

Especially, a qubit-based sensor has been considered as an attractive approach to achieve a high sensitivity where a qubit is used as a probe of the 
magnetic field sensors. 
When the frequency of the qubit is shifted due to the external magnetic fields, Ramsey type measurements let us detect the amplitude of the magnetic fields.
There are many experimental demonstrations
along this direction such as  
neutral atoms (atomic vapor cells \cite{kominis2003subfemtotesla,fernholz2008spin,wasilewski2010quantum}, cold atomic clouds \cite{vengalattore2007high,sewell2012magnetic,ockeloen2013quantum}), trapped ions \cite{baumgart2016ultrasensitive,leibfried2004toward},  superconducting 
flux qubit \cite{bal2012ultrasensitive,toida2019electron,miyanishi2019nuclear}, and nitrogen-vacancy centers in diamond \cite{wolf2015subpicotesla,taylor2008high}.

One of the ultimate goals of enhancing the sensitivity of magnetic sensors is an efficient detection 
of 
the single (electron or nuclear) spin
\cite{taylor2008high,degen2008nanoscale},
and extensive research has been conducted, including both theoretical and experimental approaches {\cite{degen2008nanoscale,taylor2008high,maze2008nanoscale,balasubramanian2008nanoscale,schaffry2011proposed,muller2014nuclear,staudacher2013nuclear,mamin2013nanoscale,ohashi2013negatively,rugar2015proton,lovchinsky2016nuclear,zhao2012sensing,shi2015single,abe2018tutorial,rugar2004single}.
Since the single spin induces magnetic fields, a sensitive magnetic field sensor provides us with the way to detect the spin.
However, the magnetic field from the single spin is typically weak, and this requires a long measurement time to compensate the low signal to noise ratio.
So  the improvement of the sensitivity is essential for the rapid and accurate spin detection.

On the other hand, 
it is known that, for the detection of  global (spatially homogeneous) magnetic fields, an entanglement can enhance the performance of the qubit-based sensors.
Especially, Greenberger-Horne-Zeilinger (GHZ) states can be 
used to outperform the standard quantum limit that classical sensors cannot surpass \cite{leibfried2004toward,giovannetti2004quantum,pezze2008mach,giovannetti2011advances,jones2009magnetic,matsuzaki2011magnetic,chin2012quantum,macieszczak2015zeno,tanaka2015proposed,dooley2016quantum,pezze2018quantum,matsuzaki2018quantum}. 
There are many experimental demonstration to generate the GHZ states with several systems
\cite{leibfried2004toward,leibfried2005creation,song2019generation,omran2019generation}.

In this paper, we propose to use an entanglement of probe spins
for the detection of the target single spin. 
We assume that the probe spins are two-level systems (qubits), and 
the frequency of the probe spins is affected by the magnetic fields from the target spin due to the dipole-dipole interactions. 
We consider using the GHZ states of the probe spins to detect the target spin.
Although there are many previous researches about global magnetic field sensors with entanglement \cite{leibfried2004toward,giovannetti2004quantum,pezze2008mach,giovannetti2011advances,jones2009magnetic,matsuzaki2011magnetic,chin2012quantum,macieszczak2015zeno,tanaka2015proposed,dooley2016quantum,pezze2018quantum,matsuzaki2018quantum}, 
the single spin detection with the entangled probe spins has not been discussed.
Importantly, due to
the dipole-dipole interaction between the target single spin and the probe spins, the entangled probes
needs to detect the spatially inhomogeneous magnetic field. Especially, magnetic fields on the probe spins far from the target spin are much weaker 
than that of 
probe spins close to the target spin. 
Actually, for the single spin detection, the probe spins far from the target spin just induce noise without contributing the increase of the signal,
and therefore the increase of the number of the probe spins does not necessarily improve the signal to noise ratio.
This is stark contrast to the global magnetic field sensors with entangled probe spins where the signal to noise ratio is monotonically increased 
with increasing the number of the probe spins \cite{leibfried2004toward,giovannetti2004quantum,pezze2008mach,giovannetti2011advances,jones2009magnetic,matsuzaki2011magnetic,chin2012quantum,macieszczak2015zeno,tanaka2015proposed,dooley2016quantum,pezze2018quantum,matsuzaki2018quantum}.

By considering these conditions, we investigate the performance of the single spin detection with entangled states
where we optimize the number of the probe spins. 
Moreover, we compare the strategy to use the entangled probe spins with other strategies to use  either a single probe spin or separable ensemble probe spins.
We find that the single spin detection with the entangled probe spins achieves a few orders 
of magnitude better sensitivity than that with the single probe spin or the separable ensemble probe spins even under the effect of realistic noise.

The rest of this paper is organized as follows. In Sec.\ II, we introduce our setup of single spin detection with the GHZ states, 
and, we explain our measurement sequence with GHZ states. In Sec.\ III, we show the analytical and numerical 
solutions of the sensitivity to demonstrate the performance of the entangled probe spins.
In Sec.\ IV, we summarize and conclude this paper.

\section{Setup and Method}
We explain the setup of our calculation. The configuration of the target spin and ensemble of probe spins is shown in the Fig.\ 1. 
There are $L$ probe spins homogeneously distributed
 with the density of $\rho$ inside a columnar substrate. 
The total Hamiltonian, which includes the dipole-dipole interaction between the target spin and the probe spins, is given by
\begin{align}
\hat{H}&=\hat{H}_{{\rm T}}+\hat{H}_{{\rm P}}+\hat{H}_{{\rm I}}\\
\hat{H}_{{\rm T}}&=\frac{\omega^{({\rm T})} }{2}\hat{\sigma}_z^{({\rm T})} \\
\hat{H}_{{\rm P}}&=\sum_{j=1}^L\frac{\omega^{({\rm P})}}{2}\hat{\sigma}_{z,j}^{({\rm P})}\\
\hat{H}_{{\rm I}}&=G\sum_{j=1}^L\frac{\vec{\hat{\sigma}}^{({\rm T})} \cdot \vec{\hat{\sigma}}_{j}^{({\rm P})}-3\left(\vec{\hat{\sigma}}^{({\rm T})} \cdot \frac{\vec{r}_j}{|\vec{r}_j|}\right)\left(\vec{\hat{\sigma}}_{j}^{({\rm P})}\cdot \frac{\vec{r}_j}{|\vec{r}_j|}\right)}{|\vec{r}_j|^3},
\end{align}
where $\omega^{({\rm T})}$ ($\omega^{({\rm P})}$) is a resonant frequency of the target (probe) spin, 
 $G$ is a constant determined by the product of the magnetic moments of the target spin and the probe spins, and 
$\vec{\hat{\sigma}}_{j}^{({\rm P})}=(\hat{\sigma}_{x,j}^{({\rm P})},\hat{\sigma}_{y,j}^{({\rm P})},\hat{\sigma}_{z,j}^{({\rm P})})$ is a 
set of the Pauli matrices of the probe spins at $\vec{r}_j=(x_j,y_j,z_j)$. 
We define the coordinate $(x,y,z)$ as shown in the  Fig.\ 1. 
Here, we assume a large detuning between the target spin and the probe spins $\omega^{({\rm P})}\gg \omega^{({\rm T})}$.
In the rotating frame defined by the unitary transformation $\exp{\left[i(\hat{H}_{{\rm T}}+\hat{H}_{{\rm P}})t\right]}$, we remove the terms oscillating with $\omega^{({\rm T})}$ and $\omega^{({\rm P})}$ (that is rotating wave approximation), and we obtain the effective Hamiltonian
\begin{align}
\hat{H}^{({\rm eff})}=\sum_{j=1}^L\frac{G}{(r_j^2+z_j^2 )^{3/2}}\left(1-\frac{3z_j^2}{r_j^2+z_j^2} \right) \hat{\sigma}_z^{({\rm T})} \hat{\sigma}_{z,j}^{({\rm P})},
\end{align}
where we set the cylindrical coordinate $r=\sqrt{x^2+y^2}$. 
Since we consider a case either the target spin is up or down, we replace $\hat{\sigma}_z^{({\rm T})}$ with a classical value $s=1$ or $-1$;
\begin{align}
\hat{H}^{({\rm eff})}=\sum_{j=1}^L \frac{\omega_j}{2} \hat{\sigma}_{z,j}^{({\rm P})},
\label{eq:effHamiltonian}
\end{align}
where $\omega_j=\frac{2G}{(r_j^2+z_j^2 )^{3/2}}\left(1-\frac{3z_j^2}{r_j^2+z_j^2} \right)s$.
The variations of the distance between the target spin and each probe spin causes the inhomogeneous magnetic fields acting on the probe spins. We will consider either a single spin, an ensemble of separable spins \cite{uesugi2017single}, or entangled spins, as the probe states. When we consider the single probe spin, the probe spin is attached at the center bottom of the columnar substrate. As an entanglement, we choose the GHZ state:
\begin{align}
\ket{{\rm GHZ}}=\frac{1}{\sqrt{2}}\left(\ket{\uparrow\cdots \uparrow}+\ket{\downarrow\cdots \downarrow}\right)
\end{align}
The time evolution of the quantum states under the effect of non-Markovian dephasing is described by the master equation 
\begin{align}
\frac{\partial \hat{\rho} (t)}{\partial t}=i[\hat{\rho} (t),\hat{H}^{({\rm eff})}]- \frac{t}{4T_2^2}\sum_{j=1}^L[\hat{\sigma}_{z,j}^{(P)},[\hat{\sigma}_{z,j}^{(P)},\hat{\rho} (t)]],
\label{eq:Mastereq}
\end{align}
where $\hat{\rho} (t)$ is the density operator at time $t$ and $T_2$ denotes the coherence time of the qubit.
The first term of the right-hand side in (\ref{eq:Mastereq}) denotes
the unitary time evolution of the Hamiltonian (\ref{eq:effHamiltonian}) and the second term 
denotes
the effect of the dephasing from the environment.
 \begin{figure}
 \centering
  \includegraphics[clip,width=4.0cm]{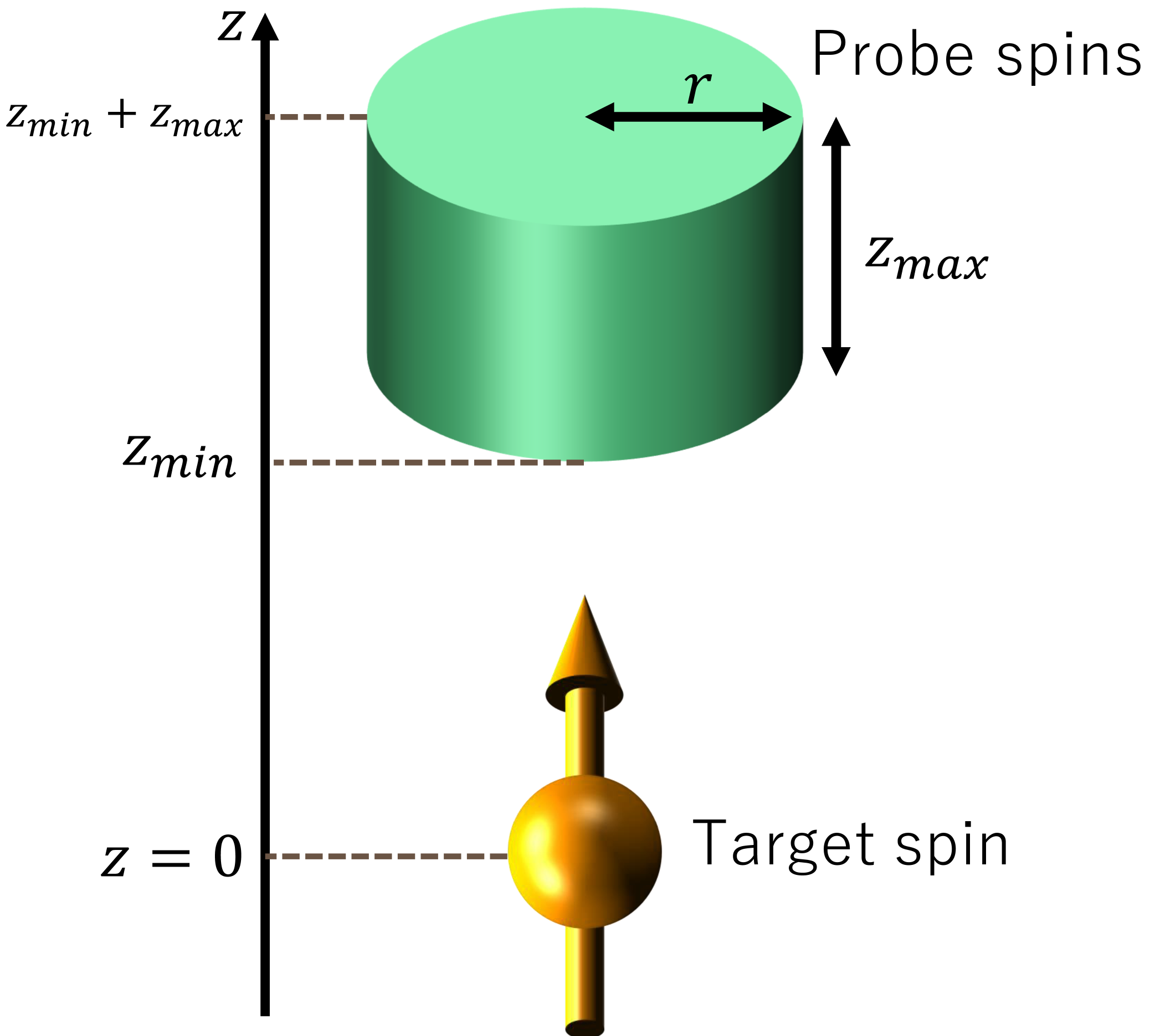}
  \caption{This illustrates the positional relation between the target spin and the probe spins.  
The target spin is located at the origin of the coordinate, and the probe spins are homogeneously distributed with the density of $\rho$ inside a columnar substrate. 
The quantization axis of the probe spins and target spin are along the $z$ axis, and this system has a rotational symmetry around the $z$ axis.}
 \end{figure}

We investigate three protocols for the single spin detection, which corresponds to the estimation of the value $s$ in this paper.
The first one is to use a single probe spin for the detection of the target spin, as shown in Table I (i).
The second one (ii) is to utilize an ensemble of separable probe spins to detect the target spin, which was theoretically studied in \cite{uesugi2017single}.
The third one (iii) is to exploit the entangled probe spins to detect the target spin. We will show that the third protocol is much more efficient than the other two.

We describe the measurement sequence. Firstly, prepare an initial state of the probe spins (where the specific form of the initial state depends on the protocol, as  
shown in Table I). 
Secondly, let the quantum state evolve by Eq.\ (\ref{eq:Mastereq}) for a time $t$. Thirdly, measure the state by a specific readout basis (which also depends on the protocol as
shown in Table I). Finally, we repeat 1-3 steps $N$ times. We assume that the state preparation time and readout time is negligibly small, and we can approximately 
obtain
 $N\simeq T/t$ where $T$ is a given total measurement time.
\begin{table}
\centering
\caption{Three protocols to estimate $s$}
  \begin{tabular}{|l||c|c|} \hline
   Probe spins  & Initital states & Readout basis  \\ \hline \hline
  (i) Single   & $\ket{+}$ & $(\ket{\uparrow}\pm i\ket{\downarrow})/\sqrt{2}$  \\
\shortstack[l]{\\ (ii) Ensemble  \\ \, \,(separable)}
 & \shortstack{$\ket{++\cdots +}$\\  \, } & \shortstack{$\otimes_j[(\ket{\uparrow}\pm\ket{\downarrow})/\sqrt{2}]$\\  \, }  \\
   (iii) Entangled & $\ket{{\rm GHZ}}$ & $(\ket{\uparrow\cdots\uparrow}\pm i\ket{\downarrow\cdots\downarrow})/\sqrt{2}$  \\ \hline
  \end{tabular}
\end{table}
\section{Results}
For the first (i) and second protocol (ii), the sensitivity of $s$  to detect the single spin has been analyzed in \cite{uesugi2017single}.
So, in this paper, we explain how to derive the sensitivity of $s$ with the third protocol (iii) where the probe spins are entangled.
In the table II, we summarize the sensitivity of these three protocols for the comparison of our entanglement protocol with the other protocols.

We describe how we can derive the analytical form of the solution of the Eq.\ (\ref{eq:Mastereq}) as follows.
$
\hat{\rho} (t) = \frac{1}{2}[\ket{\uparrow\cdots \uparrow}\bra{\uparrow\cdots \uparrow}+\ket{\downarrow\cdots \downarrow}\bra{\downarrow\cdots \downarrow}]
+\frac{e^{i \sum_j \omega_j t - L\left(\frac{t}{T_2}\right)^2}}{2}\ket{\downarrow\cdots \downarrow}\bra{\uparrow\cdots \uparrow}+h.c.
$, and we also obtain the expectation value $p=\mathrm{Tr}[\hat{P}_{{\rm meas}}\hat{\rho} (t)]=\frac{1}{2}\pm\frac{1}{2}e^{- L\left(\frac{t}{T_2}\right)^2}\sin{\left(\sum_j\omega_j t\right)}\simeq\frac{1}{2}\pm\frac{1}{2}e^{- L\left(\frac{t}{T_2}\right)^2}\left(\sum_j\omega_j t\right)$, where $\hat{P}_{{\rm meas}}$ is the measurement operator of the readout basis as shown in Table I, and we assume the weak magnetic fields $\left|\sum_j\omega_j t\right|\ll 1$.
In order to estimate the error of $s$, we calculate $\delta s:=\frac{\delta p}{\sqrt{N}\left|\frac{\partial p}{\partial s}\right|}$, where $\delta p$ ($:=\sqrt{p(1-p)}$) is the standard deviation of $p$.
To minimize the uncertainty, we choose $t=T_2/2\sqrt{L}$, and 
we obtain $\delta s=\frac{\sqrt{2}e^{1/4}}{\sqrt{TT_2}}\frac{L^{1/4}}{\left|\sum_j\frac{\partial\omega_j}{\partial s}\right|}$.
Here, we take a continuous limit $\left|\sum_j\frac{\partial\omega_j}{\partial s}\right|\simeq\rho\left|\iiint dxdydz\frac{2G}{(r_j^2+z_j^2 )^{3/2}}\left(\frac{3z_j^2}{r_j^2+z_j^2}-1 \right)\right|=4\pi G\rho\left|\frac{z_{{\rm max}}}{\sqrt{r^2+z_{{\rm max}}^2}}-\frac{z_{{\rm min}}}{\sqrt{r^2+z_{{\rm min}}^2}}\right|$. 
This approximation is justified as $r,z_{{\rm max}},z_{{\rm min}}\gg \rho^{-1/3}$, where $\rho^{-1/3}$ is the average distance among each probe spin.
Moreover, we optimize the size of the columnar substrate (the number of the probe spins).
Using $L=\rho \pi r^2 (z_{{\rm max}}-z_{{\rm min}})$, $\delta s=\frac{\sqrt{2}e^{1/4}}{4G\pi^{3/4}\sqrt{TT_2}}\frac{z_{{\rm min}}^{3/4}}{\rho^{3/4}}\times f(\tilde{r},\tilde{z}_{{\rm max}})$, where $f(\tilde{r},\tilde{z}_{{\rm max}})=[\tilde{r}^2 (\tilde{z}_{{\rm max}}-1)]^{1/4}\times \left(\frac{\tilde{z}_{{\rm max}}}{\sqrt{\tilde{r}^2+\tilde{z}_{{\rm max}}^2}}-\frac{1}{\sqrt{\tilde{r}^2+1}}\right)^{-1}$, and $\tilde{r}=r/z_{{\rm min}}, \tilde{z}_{{\rm max}}=z_{{\rm max}}/z_{{\rm min}}$.
We 
numerically minimize
$f(\tilde{r},\tilde{z}_{{\rm max}})=4.14$ as $\tilde{r}=1.87, \tilde{z}_{{\rm max}}=4.30$, and in this case, we obtain the number of the probe spins $L=\rho \pi \tilde{r}^2 (\tilde{z}_{{\rm max}}-1)z_{{\rm min}}^{3}=35.9\times \rho z_{{\rm min}}^{3}$.

We discuss our analytical results of the sensitivity. Firstly, we
 compare our result with the previous results (i) using the single probe spin  or (ii) using separable ensemble probe spins  for the single spin detection
in the table II.
Clearly, the scaling of $z_{\rm{min}}$ of our scheme ($\delta s = O(z_{\rm{min}}^{3/4})$) is different from those in the other two schemes
($\delta s = O(z_{\rm{min}}^{3})$ or $\delta s = O(z_{\rm{min}}^{3/2})$). These show that entanglement probe provides us with more efficient single spin detection than 
the other schemes
especially when
the target spin is located far from the probe spins.
Secondly, we compare our results with the sensors to measure  global magnetic fields $B$ with $L$ qubits in the table II.
It is worth mentioning that, although we firstly analyze the performance to use the entangled states for the single spin detection, previous researches show that 
entangled states can enhance the performance of
 the global magnetic field sensors\cite{leibfried2004toward,giovannetti2004quantum,pezze2008mach,giovannetti2011advances,jones2009magnetic,matsuzaki2011magnetic,chin2012quantum,macieszczak2015zeno,tanaka2015proposed,dooley2016quantum,pezze2018quantum,matsuzaki2018quantum}. 
From the table, we have found that the number of the qubits $L$ to measure the global magnetic fields corresponds to the density $\rho $ to detect the single spin.
For example, the separable ensemble probe spins provide $\delta B=O(L^{-1/2})$ or $\delta s=O(\rho ^{-1/2})$ (depending on the global magnetic field measurement or single spin detection)
 while 
the entangled probe spins provide $\delta B=O(L^{-3/4})$ or $\delta s=O(\rho ^{-3/4})$.
For the global magnetic field sensors with the entangled probe spins, we can just increase the number of the qubits to
improve the signal to noise ratio. On the other hand, for the single spin detection, the sensitivity could be improved if we could increase the density
while the other parameters were fixed. However, usually, the increase of the density degrades the coherence time $T_2$, and so it is not straightforward to increase just density
with the same parameters, which is stark contrast with the global magnetic field sensing.


\begin{table}
\centering
\caption{Summary of results}
  \begin{tabular}{|l||c|c|} \hline
 Probe spins    & Global $B$& Single spin detection  \\ \hline \hline
  \shortstack[l]{\\ (i) Single  \\  \, }& \shortstack{\\ $\delta B=O(L^0)$\\ \,} & \shortstack{\\ $\delta s^{({\rm s})}=\frac{\sqrt{2}e^{1/4}}{4G\sqrt{TT_2}}z^3_{\rm min}$\\ \cite{uesugi2017single}}  \\
\shortstack[l]{\\ (ii) Ensemble  \\ \,\,(separable)}
 & \shortstack[l]{\\ \, \\ $\delta B=O(L^{-1/2})$\\ (standard \\ quantum limit)}  & \shortstack{\\ $\delta s^{({\rm sep})}=\frac{5.32\sqrt{2}e^{1/4}}{4G\sqrt{\pi}\sqrt{TT_2}}\frac{z^{3/2}_{\rm min}}{\sqrt{\rho}}$ \\ \cite{uesugi2017single} }\\
\shortstack[l]{\\   (iii) Entangled  \\  \, }& \shortstack{\\$\delta B=O(L^{-3/4})$ \\ \cite{matsuzaki2011magnetic,chin2012quantum,tanaka2015proposed}} & \shortstack{\\$\delta s^{({\rm en})}=\frac{4.14\sqrt{2}e^{1/4}}{4G\pi^{3/4}\sqrt{TT_2}}\frac{z^{3/4}_{\rm min}}{\rho^{3/4}}$\\ (our results)}   \\ \hline
  \end{tabular}
\end{table}

 \begin{figure}
 \centering
  \includegraphics[clip,width=7.0cm]{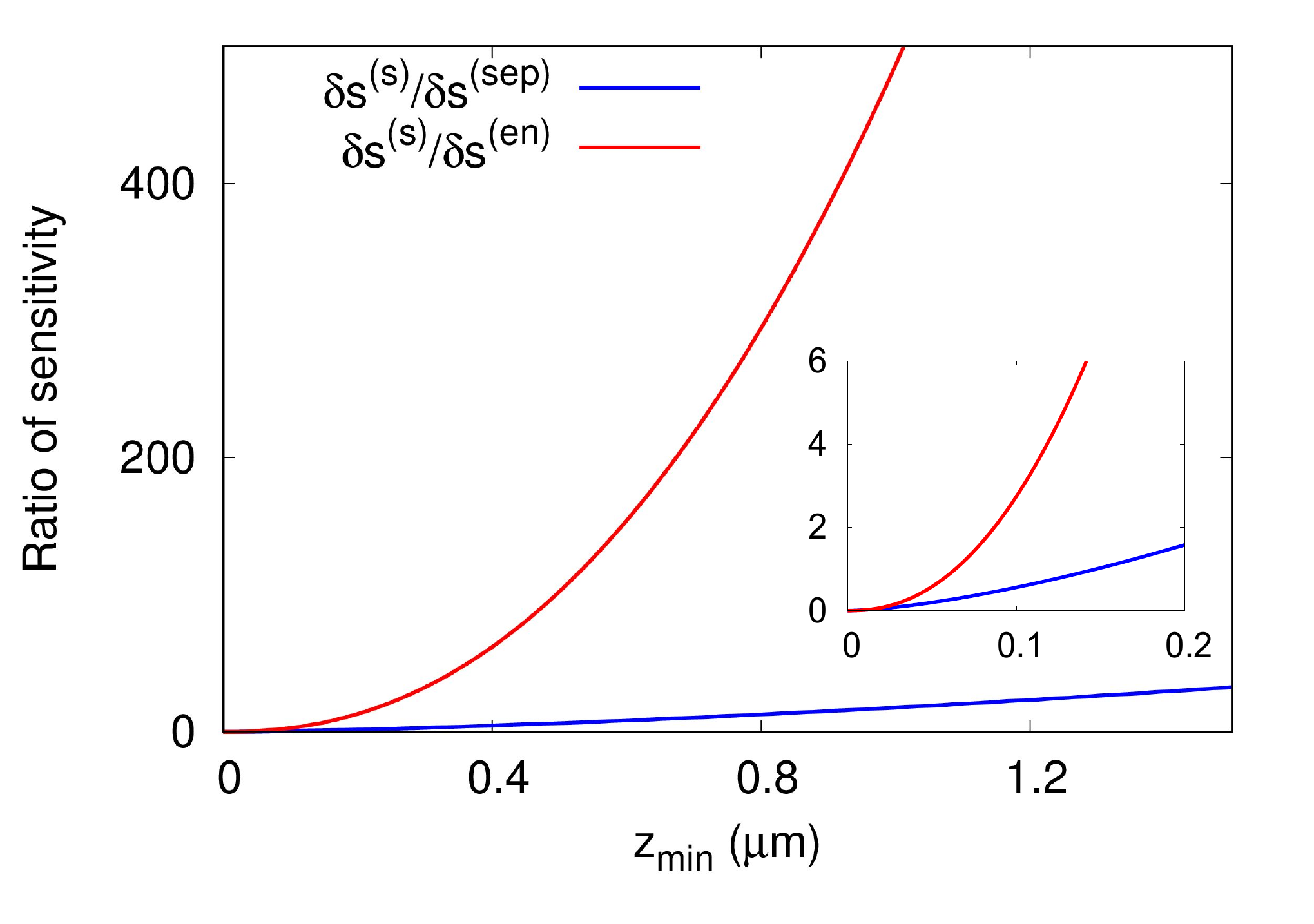}
  \caption{Plot of the ratios of sensitivities $\delta s^{({\rm s})}/\delta s^{({\rm sep})}$ and $\delta s^{({\rm s})}/\delta s^{({\rm en})}$ 
against $z_{{\rm min}}$.
Here,
$\delta s^{({\rm s})}$,
$\delta s^{({\rm sep})}$, 
and $\delta s^{({\rm en})}$ denote  the sensitivity of the single probe spin, separable ensemble probe spins, and entangled probe spins, respectively.
We choose the typical parameters of NV centers, $\rho=6.7\times 10^{16}/{\rm cm}^3$ and $T_2=84\mu$s for an ensemble of the NV centers, and
$T_2=2000\mu$s for a single NV center. The inset is a magnified view of the plot.
}
 \end{figure}

Here, we perform numerical simulations to evaluate the 
sensitivity
to detect a single electron spin with realistic parameters.
We consider the nitrogen vacancy (NV) centers as probe spins, 
which are one of the most promising systems to realize quantum enhanced sensors {\cite{wolf2015subpicotesla,degen2008nanoscale,taylor2008high,muller2014nuclear,staudacher2013nuclear,maze2008nanoscale,balasubramanian2008nanoscale,lovchinsky2016nuclear,zhao2012sensing,shi2015single,abe2018tutorial,
schaffry2011proposed,degen2008scanning,mamin2013nanoscale,rugar2015proton,ohashi2013negatively,togan2010quantum,rondin2014magnetometry,tanaka2015proposed,grezes2015storage,balasubramanian2009ultralong}.
In the Fig.\ 2, we plot $\delta s^{({\rm s})}/\delta s^{({\rm sep})}$ and $\delta s^{({\rm s})}/\delta s^{({\rm en})}$
against $z_{{\rm min}}$ where 
$\delta s^{({\rm s})}$,
$\delta s^{({\rm sep})}$, 
and $\delta s^{({\rm en})}$ denote  the sensitivity of the single probe spin, separable ensemble probe spins, and entangled probe spins, respectively.
These analytical form are shown in Table II.
Here, for the numerical simulations, 
we choose $\rho=6.7\times 10^{16}/{\rm cm}^3$ and $T_2=84\mu$s for an ensemble of the NV centers \cite{wolf2015subpicotesla,grezes2015storage}, while we consider
$T_2=2000\mu$s for a single NV center \cite{balasubramanian2009ultralong}.
From this plot, we find that the ratio of the uncertainty is $\delta s^{({\rm s})}/\delta s^{({\rm en})}\simeq 500$
 at $z_{{\rm min}}=1\mu$m, which means that the probe state with the entangled spins achieve $500$ 
times higher sensitivity than that of the single spin. 
On the other hand, for the probe state with the separable spins, the ratio is $\delta s^{({\rm s})}/\delta s^{({\rm sep})}\simeq 17$  at $z_{{\rm min}}=1\mu$m. These results show that the entangled probe spins achieves much higher sensitivity than that of the separable probe spins.
Moreover, with the entangled probe spins, we have a wider parameter ranges about $z_{\rm{min}}$ to beat the sensitivity of the single spin than with the separable probe spins.
From the Fig.\ 2 (inset), we can see that $\delta s^{({\rm s})}/\delta s^{({\rm sep})}>1$ at $z_{{\rm min}}>0.15 \mu$m, while $\delta s^{({\rm s})}/\delta s^{({\rm en})}>1$ at $z_{{\rm min}}>$0.065$\mu$m.
The optimal number $L$ of the probe spins with entangled states is proportional to $z_{{\rm min}}^3$, and
in the case of $z_{{\rm min}}=1\mu$m, it is $L\simeq2.4\times 10^6$.

Furthermore, we numerically investigate the necessary time to detect a single spin.
When the target spin is an electron spin, we can numerically calculate $\delta s $ with the parameters explained above, and $\delta s$ decreases as 
the total measurement time $T$ increases. We define the necessary time for the single spin detection as $T$ such that $\delta s=1$ is satisfied.
Also, we introduce 
$T^{({\rm s})}$, $T^{({\rm s})}$, and $T^{({\rm en})}$  to denote the necessary time for the single spin detection when the probe is a single spin, a separable ensemble,
and an entangled state, respectively. We obtain
$T^{({\rm s})}\sim 10^4$s, $T^{({\rm sep})}\sim 10^2$s, and  $T^{({\rm en})}\simeq 0.1$s
at $z_{{\rm min}}=1 \mu$m. Therefore, the entangled probe provides us with a much more rapid detection of
the single target spin.

\section{Summary and Conclusion}
In conclusion, we have shown that entangled probe spins enable us to detect the target single spin with realistic parameters much more efficiently than the single probe spin or the separable ensemble probe spins. 
We expect that by using entangled states, more efficient detection of single spin will be achieved experimentally near the future.

\begin{acknowledgments}
We are grateful to Syuhei Uesugi, S. Endo, and Junko Hayase for their assistant in this study. This work was supported by Leading Initiative for Excellent Young Researchers MEXT Japan, and is partially supported by MEXT KAKENHI (Grant No. 15H05870).
\end{acknowledgments}
\appendix



\end{document}